\documentclass[aps,showpacs,twocolumn,showkeys,prb]{revtex4}
\usepackage{graphicx,color,epsfig,rotate}
\usepackage{times,xspace}
\usepackage{amsbsy,amssymb,amsmath,bm}
\usepackage{dcolumn}% Align table columns on decimal point
\usepackage{bm}% bold math
\input{epsf.sty}

\def\bbbc{{\mathchoice {\setbox0=\hbox{$\displaystyle\rm C$}\hbox{\hbox
to0pt{\kern0.4\wd0\vrule height0.9\ht0\hss}\box0}}
{\setbox0=\hbox{$\textstyle\rm C$}\hbox{\hbox
to0pt{\kern0.4\wd0\vrule height0.9\ht0\hss}\box0}}
{\setbox0=\hbox{$\scriptstyle\rm C$}\hbox{\hbox
to0pt{\kern0.4\wd0\vrule height0.9\ht0\hss}\box0}}
{\setbox0=\hbox{$\scriptscriptstyle\rm C$}\hbox{\hbox
to0pt{\kern0.4\wd0\vrule height0.9\ht0\hss}\box0}}}}

\begin{document}
\title{Multiferroic behavior in the new double-perovskite Lu$_2$MnCoO$_6$}

\author{S. Y\'{a}\~{n}ez-Vilar$^1$, E. D. Mun$^2$, V. S. Zapf$^2$, B. G. Ueland$^3$, J. Gardner$^{4,5}$, J. D. Thompson$^3$, J. Singleton$^2$, M. S\'{a}nchez-And\'{u}jar$^1$, J. Mira$^6$, N. Biskup$^7$, M. A. Se\~{n}ar\'{i}s-Rodr\'{i}guez$^1$, C. D. Batista$^8$}

\affiliation{
$^1$Dpto. Qu\'{i}mica Fundamental U. Coru\~{n}a, 15071 A Coru\~{n}a (Spain)\\
$^2$National High Magnetic Field Laboratory (NHMFL) Materials Physics and Applications - Condensed Matter and Magnetic Science (MPA-CMMS), Los Alamos National Lab (LANL) Los Alamos NM 98545 (USA)\\
$^3$MPA-CMMS, LANL, Los Alamos, NM 87545 (USA)\\
$^4$ NIST Center for Neutron Research National Institute of Standards and Technology, 100 Bureau Drive Gaithersburg Maryland 20899 (USA)\\
$^5$Indiana University, Bloomington, Indiana 47408 (USA)\\
$^6$Dpto. F\'{i}sica Aplicada U. Santiago de Compostela 15782 Santiago de Compostela (Spain)\\
$^7$Dpto. Tecnolog\'{i}as de la Informaci\'{o}n Inst. de Ciencia de Materiales 28040 Madrid (Spain)\\
$^8$Theory division, LANL, Los Alamos, NM 87545 (USA) \\
}

\date{\today}

\begin{abstract}

We present a new member of the multiferroic oxides, Lu$_2$MnCoO$_6$, which we have investigated using X-ray diffraction, neutron diffraction, specific heat, magnetization, electric polarization, and dielectric constant measurements. This material possesses an electric polarization strongly coupled to a net magnetization below 35 K, despite the antiferromagnetic ordering of the $S = 3/2$ Mn$^{4+}$ and Co$^{2+}$ spins in an  $\uparrow \uparrow \downarrow \downarrow$ configuration along the c-direction. We discuss the magnetic order in terms of a condensation of domain boundaries between $\uparrow \uparrow$ and $\downarrow \downarrow$ ferromagnetic domains, with each domain boundary producing a net electric polarization due to spatial inversion symmetry breaking. In an applied magnetic field the domain boundaries slide, controlling the size of the net magnetization, electric polarization, and magnetoelectric coupling. 

\end{abstract}

\maketitle

\section{Introduction}

Magneto-electric (ME) multiferroics are materials with long-range electric and magnetic order. \cite{Fiebig05} Understanding how multiple order parameters coexist and couple is interesting in and of itself. However, ME multiferroics also have potential applications to magnetic storage, novel circuits, sensors, microwave and high-power applications. \cite{Scott07,Chu08} Achieving strong ME coupling between {\it net} magnetization and {\it net} electric polarization is particularly important for applications. To date however, multiferroics are rare, and those with significant ME coupling even more so. Transition magnetic oxides have been attracting the most attention in this field recently due to their relatively high magnetic ordering temperatures and tendency to form large electric polarizations. \cite{Hill00,Cheong07} Those with the strongest ME coupling have complex spin textures that break spatial-inversion symmetry (SIS) and alter the lattice so as to generate an electric polarization. \cite{Kimura03,Goto04,Hur04,Lawes05,Katsura05,Cheong07,Kenzelmann07,Arima07,Kimura07} The trouble is that many of these complex spin textures don't produce any net magnetization. 

%In the simplest transition metal oxides, magnetic behavior and ferroelectricity are inimical because the metal ions prefer to be centered in their oxygen octahedra, whereas an electric polarization requires an off-center distortion. \cite{Hill00} However, there are a number of ways to circumvent this problem and so a growing number of multiferroic magnetic oxides materials are being reported in the literature. \cite{Cheong07} 

\epsfxsize=180pt
\begin{figure}[tbp]
\epsfbox{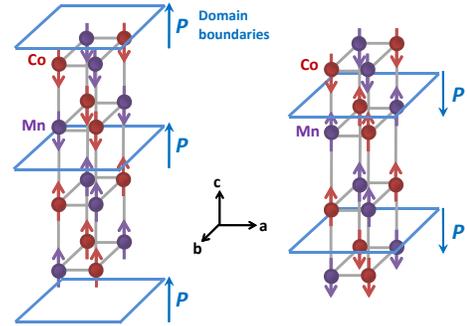}
\caption{Derived $\uparrow \uparrow \downarrow \downarrow$ Mn$^{4+}$ $S = 3/2$ and Co$^{2+}$ $S = 3/2$ spin orientations along the c-axis at $T = 4$ K. Domain boundaries refer to the boundary between $\uparrow \uparrow$ and $\downarrow \downarrow$. Two scenarios (left and right) for the location of domain boundaries are shown, along with possible resulting electric polarizations $P$.  In the a-b plane, an additional slow and incommensurate modulation of the spins occurs such that $\vec{k} = (0.0223(8), 0.0098(7), 0.5)$.}
\label{spins}
\end{figure}

Ca$_3$MnCoO$_6$ \cite{Choi08,Jo09} was recently found to exhibit a net hysteretic magnetization below 14 K, coupled to a ferroelectric polarization that is suppressed in magnetic fields of 10 T. This compound forms chains of alternating Mn$^{4+}$ $S = 3/2$ and Co$^{2+}$ $S = 1/2$ ions, with the chains in turn arranged in magnetically frustrated triangles. Magnetic exchange is mediated via oxygens that form edge-sharing octahedra around each ion. An $\uparrow \uparrow \downarrow \downarrow$ magnetic structure along Mn-Co chains breaks spatial inversion symmetry and induces electric polarization. Although an $\uparrow \uparrow \downarrow \downarrow$ spin configuration is found in zero magnetic field, a net magnetization with hysteresis is observed in applied magnetic fields.

\begin{figure}[tbp]
\epsfbox{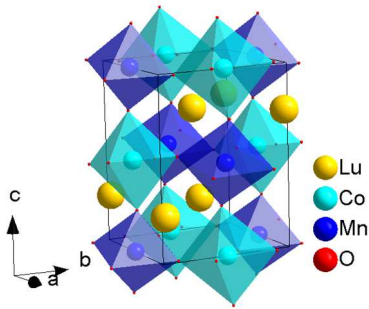}
\caption{Monoclinic crystal structure of Lu$_2$MnCoO$_6$, showing the tilted oxygen octahedra surrounding alternating Mn$^{4+}$ (dark blue) and Co$^{2+}$ (light blue) ions. Yellow Lu ions are also shown.}
\label{structure}
\end{figure}

Here we present results on a new compound, Lu$_2$MnCoO$_6$ in which we also observe $\uparrow \uparrow \downarrow \downarrow$ magnetic order along Co-Mn chains, that creates an electric polarization. We address the issue of how this antiferromagnetic structure can produce a net magnetization. Although the magnetic field-coupled electric polarization of Lu$_2$MnCoO$_6$ is smaller than in Ca$_3$MnCoO$_6$, partially due to its polycrystalline nature, the transition temperature for Lu$_2$MnCoO$_6$ is higher (35 K), and the magnetic field required to suppress electric polarization is lower (1.2 T), bringing us a step closer to useful temperatures and magnetic fields. The magnetic structure is also simpler with no frustrated triangular arrangement of the Co-Mn chains (see figures \ref{spins} and \ref{structure}), thus unraveling the physics is more straightforward. In Lu$_2$MnCoO$_6$ both Co$^{2+}$ and Mn$^{4+}$ spins are $S = 3/2$ instead of Co$^{2+}$ $S = 1/2$ and Mn$^{4+}$ $S = 3/2$, and the oxygen octahedra are corner-sharing rather than edge-sharing. These suggests that the $\uparrow \uparrow \downarrow \downarrow$ magnetic structure coupling to electric polarization can be a wide-spread mechanism for coupling of net magnetism and electric polarization and can be pushed in the direction of useful temperatures and magnetic fields for applications.

%The ordered double perovskite La$_2$NiMnO$_6$ \cite{Rogado05} has also gained attention in this field due to its high dielectric constant and large magnetocapacitive effect for the case of the ferromagnetic insulating compound. In thin films of La$_2$NiMnO$_6$ \cite{Singh07} a coupling between magnetic and dielectric properties has also been reported.

%Ab initio calculations provide a prediction of multiferroicity in the double perovskite Y$_2$NiMnO$_6$. Here an E*-type magnetic structure is predicted due to magnetic frustration in ferromagnetic-ferromagnetic-antiferromagnetic triangles. \cite{Kumar11} This magnetic order breaks SIS and thus electric polarization is predicted to be induced.  In this work, we have chosen the isostructural compound Lu$_2$MnCoO$_6$. 

\epsfxsize=150pt

\epsfxsize=200pt
\begin{figure}[tbp]
\epsfbox{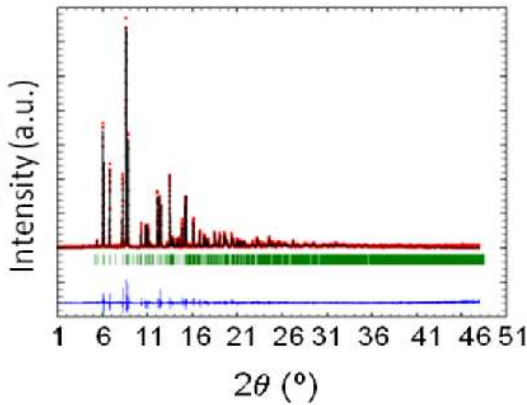}
\caption{Room temperature SXRPD patterns of Lu$_2$MnCoO$_6$ and corresponding Rietveld refinement.
Key: Observed (dots), calculated (solid line) and difference (at the bottom) profiles. The tick marks indicate the positions of the allowed Bragg reflections.}
\label{SXRPD}
\end{figure}

\section{Materials and Methods}

We synthesized a polycrystalline sample of Lu$_2$MnCoO$_6$ by a nitrate decomposition method using Lu$_2$O$_3$ (Aldrich, 99.9\%), Co(NO$_3$)$_2$6H$_2$O (Aldrich, 98\%), and Mn(NO$_3$)$_2$5H$_2$O (Aldrich, 98\%) as starting materials. We performed numerous syntheses to obtain a pure sample because there is frequently a small quantity of Lu$_2$O$_3$. The procedure was as follows: Lu$_2$O$_3$ was first converted into the corresponding nitrate by dissolution in 30\% nitric acid. This product was then added to an aqueous solution in which stoichiometric amounts of Mn(NO$_3$)$_2$ $\cdot$ H$_2$O and Co(NO$_3$)$_2 \cdot$ 6H$_2$O were also dissolved. The resulting solution was heated at $200\,^{\circ}\mathrm{C}$ until it formed a brown resin, whose organic matter was subsequently decomposed at $400\,^{\circ}\mathrm{C}$. The obtained precursor powder was then treated at $800\,^{\circ}\mathrm{C}$/60 h, $900\,^{\circ}\mathrm{C}$/24 h, $1000\,^{\circ}\mathrm{C}$/24 h, $1100\,^{\circ}\mathrm{C}$/96 h, $1150\,^{\circ}\mathrm{C}$/96 h and $1200\,^{\circ}\mathrm{C}$/48 h with intermediate gridings. The sample was then cooled at $42 \,^{\circ}\mathrm{C}$/hr to room temperature.

The purity of the material was initially checked by conventional X-ray powder diffractometry (XRPD) in a Siemens D-5000 diffractometer at room temperature using Cu K$\mathrm{\alpha}$ radiation. Additional studies were carried out with high resolution synchrotron X-ray powder diffraction (SXRPD) in the ID31 beamline ($\lambda = 0.3994 \mathrm{\AA}$) at the European Synchroton Research Facility (ESRF) in Grenoble, France. For this purpose, the samples were loaded in a borosilicate capillary ($\mathrm{\phi} = 0.3$ mm) and rotated during data collection. Rietveld refinements were performed with the Fullprof program suite. \cite{Carvajal92} The peak shapes were described by a pseudo-Voigt function, the background was modeled with a 6-term polynomial, and in the final steps of the refinement all atomic coordinates and isotropic temperature factors were included.
Iodometric titrations were carried out to analyze the oxygen content of the material. The sample was dissolved in acidified KI solutions and the I$_2$ generated was titrated against a thiosulphate solution. The whole process was carried out under an argon atmosphere. The granulometry of the sample was studied by Scanning Electron Microscopy (SEM), in a JEOL 6400 microscope.

Neutron diffraction measurements were made at the National Institute of Standards and Technology Center for Neutron Research (NCNR) on the BT1 High Resolution Powder Diffractometer.  The (311) reflection of Ge or Cu was used to produce monochromatic neutron beams with wavelengths of $\lambda = 2.079$ and $1.540 \rm{\AA}$, respectively. 15', 20', and 7' collimators were used on the in-pile, monochromated, and diffracted beams. The sample was loaded in a V can filled with He exchange gas and mounted in a closed-cycle He refrigerator capable of cooling down to T = 4 K. Data were refined using the FullProf program suite, \cite{Carvajal92} and the program k-search \cite{Carvajal92} was used to help determine the propagation vector of the magnetic order.  Representational analysis to determine the symmetry allowed magnetic structures was performed using the programs BasIreps \cite{Carvajal92} and SARAh. \cite{Wills00}  Quoted uncertainties represent one standard deviation.     

Pressed pellet samples were used for all the measurements described below.

DC magnetization measurements were made in a Quantum Design (QD) Vibration Sample Magnetometer (VSM) at the National High Magnetic Field Laboratory (NHMFL) in Los Alamos, NM in magnetic fields up to 13 T, with a DSM 1660 VSM in Spain, and with an extraction magnetometer \cite{Detwiler00} in a "short pulse" magnet (7 ms rise time, 100 ms total pulse time) up to 60 T at the NHMFL. AC magnetometry was measured in a QD AC superconducting quantum interference device (SQUID) for frequencies between 10 and 1000 Hz in an applied oscillating magnetic field of 3 x 10$^{-4}$ T. 

Specific heat $C$ was measured by the relaxation method in a QD Physical Properties Measurement System (PPMS) for temperatures down to 2 K and magnetic fields up to 13 T. 

The complex dielectric permittivity was measured with a precision LCR-meter Quadtech model 1920 over the frequency and temperature range $20\,\mathrm{Hz} \leq f \leq 106\,\mathrm{Hz}$ and $10\,\mathrm{K} \leq T \leq 300\,\mathrm{K}$. 

Dielectric measurements in magnetic fields up to 14 T were performed at various temperatures for frequencies between 10 kHz and 1 MHz. The sample used for these measurements had an area of 26 mm$^2$ and a thickness of 0.8 mm. Gold was deposited on the surfaces to ensure good electrical contact. 

Electric polarization $P$ as a function of magnetic field $H$ was measured in pulsed magnetic fields \cite{Zapf10} for $\vec{P}$ parallel and perpendicular to $\vec{H}$. Platinum contacts were sputtered onto the samples with a cross-section area of 4 mm$^2$ and a thickness of 0.1 mm. The measured quantity is the magnetoelectric current d$P$/d$t$, generated as charges are drawn from ground onto the sample contacts to screen the sample's changing electric polarization during the magnetic field pulse. d$P$/d$t$ was measured using a Stanford Research 570 current to voltage amplifier and then integrated to find $\Delta P(H) = P(H) - P(H=0)$.

\begin{table}
\caption {Structural parameters after the Rietveld refinement of the SXRPD pattern with a monoclinic symmetry (S.G: P2$_1$/n) at room temperature. The estimated errors are in parentheses.}
\label{table1}
	\centering
	\begin{tabular}{|l|l|l|l|}
  \hline
  \multicolumn{4}{|c|}{$a = 5.1638(1) \mathrm{\AA}$, $b = 5.5467(1) \mathrm{\AA}$, $c = 7.4153(1) \mathrm{\AA}$} \\
  \multicolumn{4}{|c|}{$\beta = 89.665(1)$} \\
  \hline
  Atom & x & y & z \\
  Lu & 0.5208(1) &	0.5787(1) &	0.2499(1) \\
  Co &	0	& 0.5 &	0\\
  Mn	&	0.5 &	0 &	0\\
  O1	&	0.3841(16)	& 0.9585(17)	& 0.2411(16) \\
  O2	& 0.1971(20)	& 0.1957(25)	& -0.0575(15) \\
  O3	& 0.3228(18)	& 0.6953(21)	& -0.0593(14) \\
  \hline
   \multicolumn{4}{|c|}{$R_{wp} = 14.8$   $R_p =  8.05$  $\chi^2= 1.87$} \\
  \hline

\end{tabular}
\end{table}

\begin{table}
\caption {Mn-O and Co-O bond distances and Mn-O-Co angles obtained from the room temperature refinement. Valences determined from the Bond Valence Sum (BVS) method for Mn and Co atoms are also shown. The estimated errors are in parentheses.}
\label{table2}
	\centering
	\begin{tabular}{|l|l|l|l|}
  \hline
  \multicolumn{4}{|c|}{BVS: Co valence +2.38, Mn valence +3.61} \\
  \hline  
  \multicolumn{2}{|c|}{distances (\AA)} & \multicolumn{2}{|c|}{angles (deg)} \\
  \hline
  Co-O(1) &	2.026(12) & Mn-O(1)-Co	& 141.8(5) \\
  Co-O(2)	& 2.014(15) & Mn-O(2)-Co	& 145.4(6) \\ 
  Co-O(3)	& 2.033(10) & Mn-O(3)-Co	& 142.9(4) \\
  Mn-O(1)	& 1.897(12) & & \\
  Mn-O(2)	& 1.955(13) & &\\
  Mn-O(3)	& 1.974(11) & & \\
  \hline
\end{tabular}
\end{table}

\section{Results}

\subsection{Crystal structure from X-ray diffraction}
Both neutron and X-ray diffraction measurements show that this sample is single phase and can be indexed in the monoclinic space group P2$_1$/n (see figure \ref{structure}).  The results of the iodometric titrations indicate that the sample has a very small oxygen deficiency ($\delta$) of 0.02. Scanning electron micrographs show that the morphology and microstructure of the sample consists of sintered particles with an average diameter $\phi \sim 2 \mu$m. 
The room temperature SXRPD pattern along with its refinement are shown in figure \ref{SXRPD}. Following the structure determined for La$_2$MnCoO$_6$ \cite{Daas08} and a model proposed for Y$_2$MnCoO$_6$, \cite{Troyanchuk00}, the constraint of complete transition metal cationic ordering was imposed to this refinement (Wyckoff positions 2c and 2b sites for the Mn and Co cations, respectively). However, as shown in the next section, our neutron diffraction data indicate that ~9\% mixing occurs between the sites.

The cell parameters, atomic coordinates, interatomic distances and metal-O-metal angles derived from the X-ray diffraction pattern are summarized in tables \ref{table1} and \ref{table2}. These values agree with those obtained from neutron scattering. From table \ref{table1} we see that the monoclinic angle $\beta$ is 89.665(1)$^{\circ}$, indicating a nearly orthorhombic structure. The Mn and Co cations are localized in corner-sharing octahedral environments with three different Mn-O and Co-O distances, listed in table \ref{table2}. The Co-O distances range from 2.026 to 2.033 $\mathrm{\AA}$, indicating that the valence for the Co ions is likely 2+. The Mn-O distances range from 1.897 to 1.974 $\mathrm{\AA}$ in the Mn-O octahedron, suggesting the presence of Mn$^{4+}$ as expected by analogy with La$_2$MnCoO$_6$.\cite{Daas08} In addition, the charges of these two cations have been estimated using the Bond Valence Sum (BVS) method. \cite{Brown85,Adams01} The calculated formal valences for Mn and Co are +3.61 and +2.38 respectively, near the expected values of Mn$^{4+}$ and Co$^{2+}$ for the fully ordered structure. In table \ref{table2} we observe that the smaller radius of Lu compared to La in this structure \cite{Rogado05} decreases the Co-O-Mn angles and thereby increases the octahedral distortions, which in turn likely reduces the effective magnetic interactions between the Co and Mn. The Lu-O distances are also shorter than the La-O distances. These results may explain the lower $T_c$ of 43 K in Lu$_2$MnCoO$_6$ compared to the $T_c$ of 280 K in La$_2$MnCoO$_6$.

\epsfxsize=350pt
\begin{figure*}[tbp]
\epsfbox{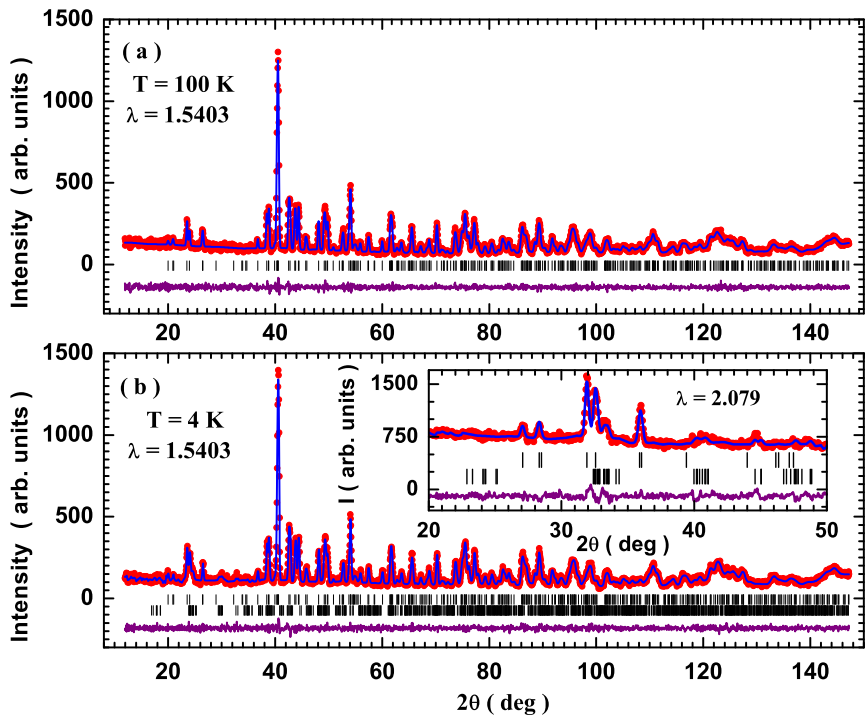}
\caption{Elastic neutron diffraction data for polycrystalline Lu$_2$MnCoO$_6$ at 100 K (a) and 4 K (b). The main panels show data taken with $\lambda = 1.540 \mathrm{\AA}$ neutrons, while the inset to figure \ref{neutrons}b shows data taken with $\lambda = 2.079 \mathrm{\AA}$ neutrons.  Red circles are experimental data, and the blue lines are fits to the data from Rietveld refinements.  Ticks underneath the data indicate symmetry-allowed Bragg positions, and purple lines beneath the ticks show the differences between the data and fits.}
\label{neutrons}
\end{figure*}

\subsection{Magnetic structure from powder neutron diffraction}

Neutron diffraction data taken at $T = 100$ and 4 K in zero magnetic field are shown in figure \ref{neutrons}a and b, respectively. 
Data at 100 K correspond to the crystal structure of the lattice and yield lattice parameters similar to those determined from the X-ray diffraction results presented above.  However, the difference in the neutron scattering lengths for Co and Mn allows us to determine that the 2c sites are occupied by 91(2)\% Co and 9(2)\% Mn, and that the 2d site are occupied by 94(2)\% Mn and 6(2)\% Co.  The "goodness of fit" indicators for figure \ref{neutrons}a are $R_{wp} = $7.46\% and $\chi^{2} = 0.75$.
Figure \ref{neutrons}b shows data at $T = 4$ K containing Bragg peaks from both the crystal structure and magnetic order.  We determined the magnetic order from the $\lambda = 2.079 \mathrm{\AA}$ data, part of which is shown in the inset to figure \ref{neutrons}b, since the higher wavelength neutrons provide greater resolution at lower values of momentum transfer $Q$.  In figure \ref{neutrons}b we include the $\lambda = 1.540 \mathrm{\AA}$ data and its refinement for easy comparison to figure \ref{neutrons}a. After an exhaustive search we determined $\vec{k} = (0.0223(8), 0.0098(7), 0.5) $ as the propagation vector of the AFM order.  This vector is only slightly incommensurate in the $a$ and $b$ directions, but the incommensurability is necessary to fit all of the magnetic peaks.  For example, the magnetic peak shown in the inset to figure \ref{neutrons}b at 33.5$^{\circ}$ cannot be fit without allowing $\vec{k}$ to be incommensurate in both the $a$ and $b$ directions.  The derived magnetic structure is shown in figure \ref{spins} and consists of an $\uparrow \uparrow \downarrow \downarrow$ type magnetic order with magnetic moments of 2.56(7) $\mu_B$/Co and 2.56(7) $\mu_B$/Mn pointed along the c-axis. We note that the moments for the Co and Mn ions were not constrained to be equal during the refinement.  The "goodness of fit" indicators for the inset to figure \ref{neutrons}b are $R_{wp} = 4.65$\% and $\chi^{2} = 1.83$.

\subsection{Thermodynamic measurements}

\epsfxsize=180pt
\begin{figure}[tbp]
\epsfbox{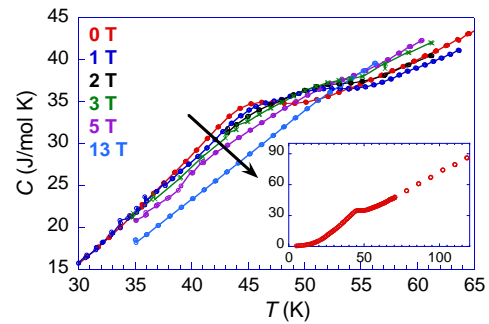}
\caption{Specific heat $C$ vs temperature $T$ at various magnetic fields between 0 and 13 T for Lu$_2$MnCoO$_6$ showing a magnetic ordering peak that broadens and evolves to higher temperatures in applied magnetic fields.}
\label{SpecificHeat}
\end{figure}

\epsfxsize=200pt
\begin{figure}[tbp]
\epsfbox{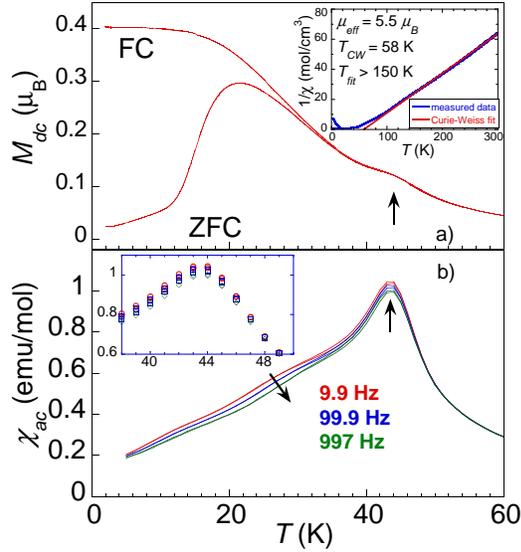}
\caption{a) DC magnetization $M_{dc}$ vs temperature $T$ measured on warming in a 0.1 T magnetic field after zero magnetic field cooling (ZFC) or magnetic field cooling (FC) from room temperature. A kink near 43 K (marked by an arrow) indicates the magnetic ordering transition. The inset shows the inverse magnetic susceptibility $1/\chi(T)$ (magnetic-field-cooled, red), which is fit by the Curie-Weiss relation (straight blue line) for $T > 150$ K. This fit yields a Curie-Weiss temperature of 58 K and a magnetic moment of 5.5 $\mu_B$/formula unit. b) AC susceptibility $\chi_{ac}$ vs temperature $T$ at frequencies of $\sim$10, 100, and 1000 Hz in an applied oscillating magnetic field of 3x10$^{-4}$ T. The ordering peak near 43 K is independent of frequency, indicating long-range order.  (1 emu = 10$^-3$ A m$^2$) }
\label{MvsT}
\end{figure}

\epsfxsize=200pt
\begin{figure}[tbp]
\epsfbox{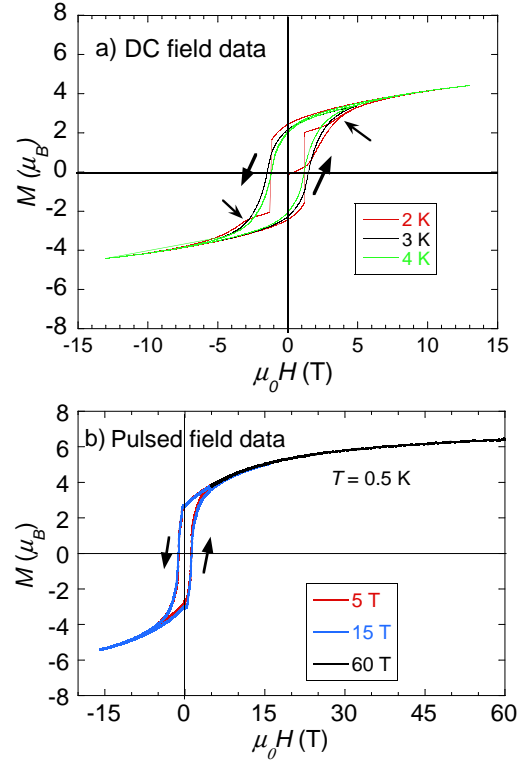}
\caption{Magnetic hysteresis loops measured in dc magnetic fields (a) and pulsed magnetic fields (b) with maximum pulses of 5, 15, and 60 T (see text). Thick arrows show the direction of the data. A coercive magnetic field of 1.21 T is observed in both data sets, and a magnetic moment of $\sim$ 6 $\mu_B$ is achieved by 0.5 K and 60 T, which would be expected from the combined Mn$^{4+}$ $S = 3/2$ and Co$^{2+}$ $S = 3/2$ moments (neglecting orbital effects). Thin arrows indicate a slight plateau in the magnetization.}
\label{MvsH}
\end{figure}

\epsfxsize=200pt
\begin{figure}[tbp]
\epsfbox{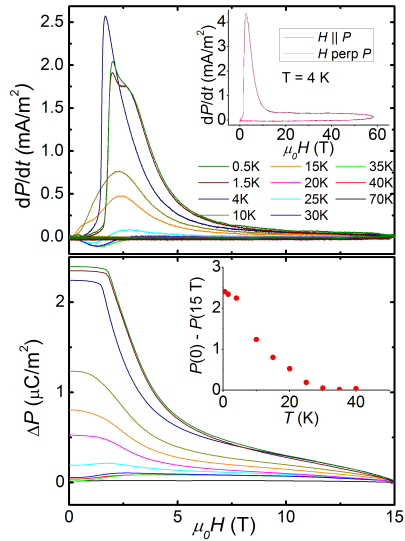}
\caption{a) Measured change in electric polarization with time, d$P$/d$t$ as a function of magnetic field $H$ for various temperature $T$ during a rapid magnetic field pulse for the geometry $\vec{P}  || \vec{H}$. Before measuring d$P$/d$t$, the sample was poled by applying an electric field of 2 MV/m in zero magnetic field while cooling from 70 K to the intended measuring temperature, at which point the electric field was removed and both sides of the sample were shorted. Inset shows data up to 60 T for $\vec{P}$ parallel and perpendicular to $\vec{H}$ at 4 K, with a 2 MV/m poling voltage. b) $\Delta P(H)$ determined by integrating the data in a). The inset shows $P (H = 0 \rm{T}) - P (H = 15 \rm{T})$ as a function of temperature. }
\label{PvsH}
\end{figure}

\epsfxsize=200pt
\begin{figure}[tbp]
\epsfbox{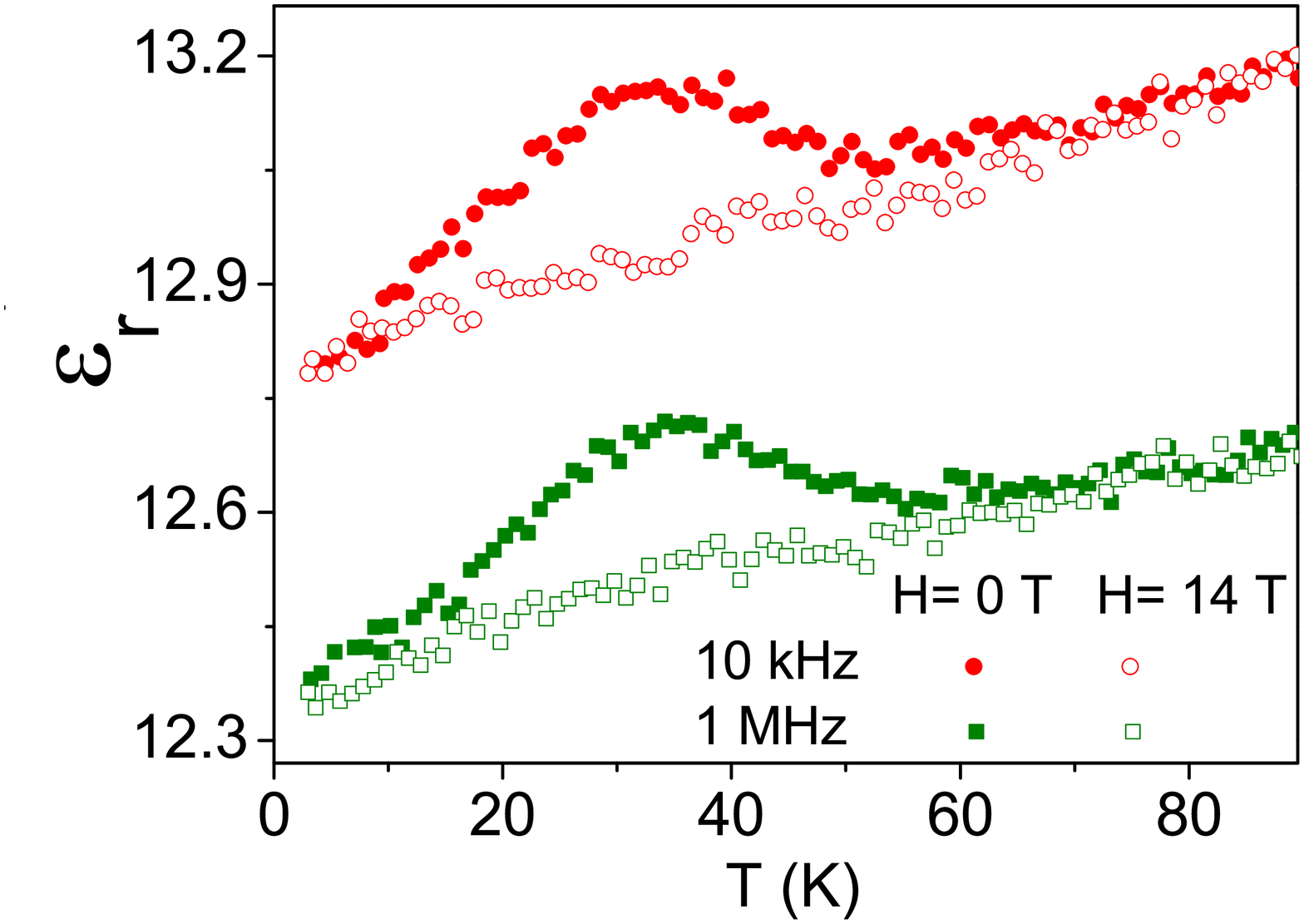}
\caption{Influence of an external magnetic field ($(\mu_{\circ}H = 14$ T) on the temperature dependence of the dielectric constant, $\epsilon_r(T)$, measured at frequencies  of 10 kHz and 1 MHz.}
\label{Dielectric}
\end{figure}

The specific heat data in figure \ref{SpecificHeat} shows a peak consistent with the onset of magnetic order below $\sim 43$ K in a polycrystalline sample. In magnetic fields up to 13 T, this peak broadens and shifts to higher temperature. This data is the total specific heat including magnetic and phonon contributions, which could not be easily subtracted.

Figure \ref{MvsT}a shows the DC magnetization vs temperature $M(T)$ measured on warming in a 0.1 T, after either magnetic field cooling (FC) in an 0.1 T magnetic field or zero magnetic field cooling (ZFC) from room temperature. A kink is observed in the magnetization near 43 K and the ZFC and FC curves separate below $\sim$35 K with the ZFC curve peaking at 20 K and then dropping to zero. The inset to figure \ref{MvsT}a shows the inverse susceptibility vs temperature with a fit to the Curie-Weiss law above 150 K. The fit results in a Curie-Weiss temperature of 58 K and an effective moment of 5.5 $\mu_B$/formula unit, which is roughly consistent with one $S = 3/2$ Co$^{2+}$ and one $S = 3/2$ Mn$^{4+}$ spin per formula unit. AC susceptibility $\chi_{ac}$ data taken at $\sim$ 10, 100, and 1,000 Hz as a function of temperature is shown in figure \ref{MvsT}b. $\chi_{ac}(T)$ shows a frequency-independent peak (within the resolution of the experiment) at 43.5 K indicating that a transition to long range magnetic order occurs.  Below 35 K, the ac susceptibility shows a small frequency-dependence indicative of slow spin dynamics. The onset of the frequency dependence arises at the same temperature below which a bifurcation between the ZFC and FC $M(T)$ curves occurs.  Though not shown here, the ZFC magnetization relaxes in the direction of the FC magnetization with a time constant of a few hours.  The observed slow spin dynamics are reminiscent of spin glass type behavior occurring below 35 K.

Magnetization vs magnetic field $M(H)$ hysteresis curves are shown in figure \ref{MvsH}a, at 2, 3, and 4 K for magnetic fields up to 13 T. At 2 K, ferromagnetic-like hysteresis is observed, with a very sudden switching of the magnetization occurring at a coercive magnetic field of 1.21 T. A plateau-like feature is seen between 1.2 and 3 T, as indicated with arrows. The switching behavior of the magnetization broadens in $H$ for the data at 3 and 4 K and the plateau disappears.  The magnetization does not fully saturate by 13 T; a moment 4.5 $\mu_B$ is achieved at 2 K and 13 T. Hysteresis curves to higher magnetic fields were measured in pulsed magnetic fields up to 60 T at the NHMFL, as shown in figure \ref{MvsH}b. This pulsed magnetic field data shows that the expected full moment of $\sim 6\mu_B$/formula unit is achieved by 60 T and 0.5 K. The identical coercive magnetic field of 1.21 T is obtained, although on the fast time scales of these pulsed magnetic field data, the reversal of the magnetization appears broader. The data shown is a combined plot of measurements from pulses with peak magnetic fields of 5, 15, and 60 T. Since the measured quantity in the extraction coil magnetometer is d$M$/d$t$, the sudden magnetization reversal at 1.21 T results in a very large d$M$/d$t$ signal that saturates the data acquisition system for the 60 T pulse. However, it is not useful to reduce the amplification or use a smaller sample in the 60 T pulse because a high sensitivity is needed to precisely measure the data near the 60 T peak magnetic field, where d$H$/d$t$ is smaller and $M(H)$ is also saturating. Instead, we measured the magnetization reversal with smaller d$H$/d$t$ pulses by reducing the peak magnetic field to 5 and 15 T. In figure \ref{MvsH}b, data for pulses with 5, 15, and 60 T peak magnetic fields are shown superimposed, with the data from the 60 T pulse only shown between 5 and 60 T. 

Our semiconducting, polycrystalline samples of Lu$_2$MnCoO$_6$ are slightly conductive at room temperature, making electric polarization and dielectric constant measurements difficult. However, with decreasing temperature the conductance decreases, reaching less than 0.1 pS below 100 K as measured with an Anderleen-Hagerlin capacitance bridge.

The change in electric polarization with magnetic field $\Delta P(H)$ was measured in pulsed magnetic fields up to 60 T after electrically poling the sample by first cooling the sample from 70 to 4 K in an electric field and then removing the electric field and shorting the two sides of the sample before measuring. Poling electric fields of 2 MV/m were used for the data shown, and $\Delta P(H)$ was found to be linear for poling electric fields between 0 and 2.5 MV/m. The measured signal, d$P(H)$/d$t$, and the integrated $\Delta P(H)$ are shown in figure \ref{PvsH}a,b. The measured $\Delta P$ is constant for magnetic fields between 0 and 1.6 T (2.6 T below 1.5 K), then drops suddenly and continues to drop at a slow and continuous rate up to 60 T (see inset). On the downsweep of the magnetic field and on subsequent $\Delta P(H)$ measurements we observe almost no $H$-dependence (the second shot after poling shows 2\% of the original $\Delta P(H)$, and subsequent shots show no resolvable $\Delta P(H)$). A significant $\Delta P(H)$ can only be observed again after re-poling. We interpret this as a magnetic field-induced \emph{suppression} of most of the electric polarization. $\Delta P(H)$ was measured for both $\vec{P}$ parallel and perpendicular to $\vec{H}$ and the same results were found in these polycrystalline samples. Data for both magnetic field directions at 4 K and up to 60 T are shown in the inset to figure \ref{PvsH}a. All the rest of the data shown was measured with $\vec{P}  || \vec{H}$.  The inset to figure \ref{PvsH}b shows the temperature dependence of $\Delta P$ between $\mu_{\circ}H = 0$ and 15 T. The onset of $\Delta P(H)$ occurs around 30 K. 

The dielectric constant as a function of temperature and magnetic field is shown in figure \ref{Dielectric}a for frequencies of 10 kHz and 1 MHz. It exhibits a broad peak near 35 K, which is the same temperature below which splitting between the ZFC and FC magnetization curves arises, frequency dependence of the ac susceptibility occurs, and $\Delta P(H)$ becomes nonzero. The peak in the dielectric constant is completely suppressed in an applied magnetic field of 14 T. 

\section{Discussion}

We interpret our results as follows: below 43 K, long-range magnetic order sets in, observed as a significant kink in the magnetization and a peak in the specific heat. Below 35 K, an electric polarization can be induced by poling in an electric field, and glassy magnetic dynamics and a hysteretic magnetization also occur. Consistent with this picture, a peak in the dielectric constant appears near 35 K (see figure \ref{Dielectric}). Neutron diffraction data at 4 K and $\mu_{\circ}H = 0$ identify a ferroelectric $\uparrow \uparrow \downarrow \downarrow$ configuration of spins along chains of alternating $S = 3/2$ Mn$^{4+}$ and $S = 3/2$ Co$^{2+}$ spins in the c-axis (see figure \ref{spins}). This spin configuration is likely the result of frustration between nearest-neighbor and next-nearest neighbor magnetic exchange interactions with opposite sign, similar to Ca$_3$MnCoO$_6$ \cite{Choi08,Jo09}.

In the following we use the term "domain boundary" to refer to the boundary between $\uparrow \uparrow$ and $\downarrow \downarrow$ spins along the c-axis. Since there are two types of ions (Co$^{2+}$ and Mn$^{4+}$), there are also two types of domain walls: the ones that are centered on a Co$^{2+}$-Mn$^{4+}$ bond and the ones centered on a  Mn$^{4+}$-Co$^{2+}$ bond. These different domain walls carry opposite electric polarizations because they break the local spatial inversion symmetry in opposite ways. In other words, the 
ferromagnetic domains walls carry an internal degree of freedom of electric polarization due to the small structural distortions
caused by the magnetostriction effects induced by the wall. This leads to the coupling between magnetism and ferroelectricity. 
In particular, a perfect $\uparrow \uparrow \downarrow \downarrow$ phase can be thought of as a condensation of domain walls whose electric polarizations are all aligned. If the sample is cooled through its transition in an electric field, it stores a net electric polarization by inducing more domains walls with one polarization than with the opposite. This electric polarization is mostly destroyed in applied magnetic fields above 1.5 T, with an additional small electric polarization persisting to 60 T. Once destroyed, the sample must be re-poled (cooled again through $T_c$ in an electric field) to regenerate the maximum electric polarization. The dielectric constant measurements also confirm the strongly magnetic field-dependent nature of the electric polarization, with the peak near 35 K completely suppressed in applied magnetic fields of 14 T.

While the  $\uparrow \uparrow \downarrow \downarrow$ spin configuration does not produce a net magnetization, we suggest that in applied magnetic fields the domain walls slide apart due to the close proximity to a ferromagnetic instability. As the domain walls become less dense, the electric polarization is also suppressed. Commensurate configurations such as $\uparrow \uparrow \uparrow \uparrow \downarrow \downarrow$ might lock in, resulting in plateaus in the magnetization. One plateau is observed near 1/3 saturation magnetization at 2 K. Magnetization data on single crystals as well as neutron diffraction data in magnetic fields are needed to test this scenario.

We note that the condensation of domain walls in the $\uparrow \uparrow \downarrow \downarrow$ leads to infinitely small domain walls and the domains themselves are as small as 7 Angstroms (c-axis lattice parameter). By contrast, conventional domain walls induced by dipole-dipole interaction in ferromagnets and multiferroics can be tens to hundreds of nm wide with domains that can be up to hundreds of mm wide. In Lu$_2$MnCoO$_6$, the condensation of domain walls that leads to the $\uparrow \uparrow \downarrow \downarrow$ configuration likely results from frustration between nearest and next-nearest neighbor interactions. Consequently, in comparison to conventional ferromagnets, the domain walls in Lu$_2$MnCoO$_6$ are far smaller and also more mobile due to the proximity to a ferromagnetic instability. This increased mobility may account for the spin glass-like frequency-dependence of the ac susceptibility below 35 K. An alternate explanation for the net hysteretic magnetization in magnetic fields is that spins tilt out of the c-axis. However this is less likely to fully explain the hysteresis and slow relaxation of the magnetization.

Although the coercive magnetic field for switching the magnetization is 1.21 T and most of the electric polarization is destroyed at 1.6 T, saturation magnetization is not reached until $\sim 60$ T, and the electric polarization continues to show a small net contribution up to this magnetic field. A likely explanation is that the 9\% Mn-Co site interchange determined from the neutron scattering data results in some Mn-Mn and Co-Co nearest neighbor pairs. In related compounds, Co-Co and Mn-Mn nearest neighbor superexchange interactions are antiferromagnetic, thus they would locally pin the domain boundaries between "up" and "down" regions of spins. 

Finally we should mention that La$_2$MnCoO$_6$ is another close relative of Lu$_2$MnCoO$_6$ that has been studied since the 1950s. In this material, confusion reigned for a long time \cite{Daas08} due to the presence of multiple phases with different Mn and Co valences, as well as Mn-Co site interchange. These problems resulted in different magnetic ordering temperatures, saturated moments, and different degrees of thermoelectric power. These structural problems mostly ensued when the oxygen deficiency $\delta$ was greater than 0.02, allowing Co$^{3+}$ and Mn$^{3+}$ to form, as well as from Mn-Co site interchange. In the case of our Lu$_2$MnCoO$_6$, we see only one magnetic phase and iodometric titrations indicate that $\delta \sim 0.02$. We do however see Co-Mn site interchange of about 9\%, which could create local antiferromagnetic interactions as discussed. 

\section{Conclusion}

In summary, Lu$_2$MnCoO$_6$ is a new member of the multiferroic oxides, showing magnetic order below 43 K, and ferroelectricity below 35 K that is strongly coupled to a net magnetism. An $\uparrow \uparrow \downarrow \downarrow$ arrangement of the spins in zero magnetic field breaks spatial inversion symmetry and induces electric polarization. We suggest that the domain walls between $\uparrow \uparrow$ and $\downarrow \downarrow$ regions slide in an applied magnetic field due to close proximity to a ferromagnetic instability, resulting in net ferromagnetic-like magnetization with a coercive field of 1.21 T that switches between states of approximately 1/3 saturation magnetization. A magnetization of $\sim$6 $\mu_B$/formula unit is eventually reached by 60 T consistent with the $S = 3/2$ spin for both Co$^{2+}$ and Mn$^{4+}$ ions. The electric polarization is strongly suppressed in magnetic fields above 1.6 T and the magnetic field-induced polarization change is $\sim 2 \mu \mathrm{C/m}^2$.

\begin{acknowledgements}

Work at the NHMFL was supported by the U.S. National Science Foundation through Cooperative Grant No. DMR901624, the State of Florida, and the U.S. Department of Energy. Measurements at LANL were also supported by the Dept of Energy's Laboratory Directed Research and Development program under 20100043DR. Work in Spain was supported by Ministerio de Ciencia e Innovaci\'{o}n MICINN (Spain) and the European Union under project FEDER MAT 2010-21342-C02. We wish to thank the European Synchrotron Radiation Facility for provision of synchrotron radiation facilities, and M. Brunelli for his assistance in using beamline ID31. We wish to thank NCNR for providing neutron scattering facilities and Mark Green for his valuable assistance in collecting data. The identification of any commercial product or trade name does not imply endorsement or recommendation by the National Institute of Standards and Technology.
\end{acknowledgements}

%\bibliography{Lu2MnCoO6}

\end{document}